# FIT: A Fog Computing Device for Speech Tele-Treatments

*Admir Monteiro[1], *Harishchandra Dubey[1], Leslie Mahler[2], Qing Yang[1], and #Kunal Mankodiya[1]

[1]Department of Electrical, Computer, and Biomedical Engineering
[2]Department of Communicative Disorders
University of Rhode Island, Kingston, RI-02881, USA
*Authors contributed equally (#Corresponding Email: kunalm@uri.edu)

*Abstract—* There is an increasing demand for smart fog-computing gateways as the size of cloud data is growing. This paper presents a Fog computing interface (FIT) for processing clinical speech data. FIT builds upon our previous work on EchoWear, a wearable technology that validated the use of smartwatches for collecting clinical speech data from patients with Parkinson's disease (PD). The fog interface is a low-power embedded system that acts as a smart interface between the smartwatch and the cloud. It collects, stores, and processes the speech data before sending speech features to secure cloud storage. We developed and validated a working prototype of FIT that enabled remote processing of clinical speech data to get speech clinical features such as loudness, short-time energy, zero-crossing rate, and spectral centroid. We used speech data from six patients with PD in their homes for validating FIT. Our results showed the efficacy of FIT as a Fog interface to translate the clinical speech processing chain (CLIP) from a cloud-based backend to a fog-based smart gateway.

## INTRODUCTION

The rising number of sensor nodes and their sampling frequency in telemedicine applications has significantly increased the size of big data in the cloud. For example, typical sensors such as smartwatches, ECG t-shirts, and sensor-rich smart homes will generate huge amounts of data surpassing the terabyte level. These applications not only increase the data size but also the communication bandwidth and computational burden on the cloud.

Fog computing that uses embedded systems holds great promise to reduce the burden of the complex medical big data by acting as a local gateway (close to the sensor nodes) for communication, computation, storage, networking, and other functions [7, 8]. We designed a fog-computing interface, FIT, that enables communication between a smartwatch and the cloud for the specific purpose of analyzing acoustic data of individuals with speech disorders.

In the current study, the smartwatch was given to patients with Parkinson's disease (PD), a neurodegenerative disorder, affecting 4 million individuals worldwide [1]. Perceptual speech impairments include reduced loudness, harsh or breathy voice quality, monotonous pitch, irregular speech rate, and imprecise articulation that can interfere with speech intelligibility [2]. One of the major challenges for speech-language pathologists (SLPs) in speech treatments is to monitor the patients when they perform homework exercises at home. FIT uses a smartwatch to collect speech data from patients when they do their

# The presented work is supported by a grant (No: 20144261) from Rhode Island Foundation Medical Research.

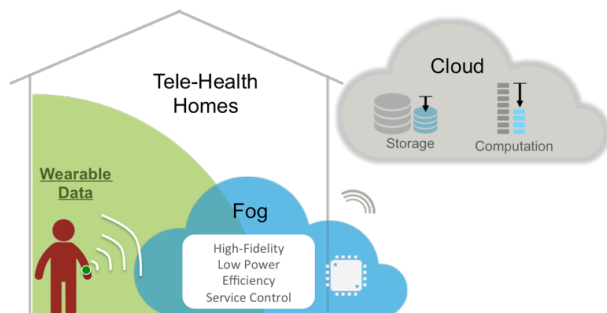

Figure 1: FIT, a Fog-driven wearable internet-of-things framework.

exercises at home and outside the clinical settings. FIT receives data from the smartwatch and performs on-demand processing, analysis and storage. The essential component of the Fog interface is Intel Edison, i.e., an embedded system. FIT made following novel contributions:

- *Embedded signal processing:* It is capable of handling psychoacoustic analysis for extracting clinical features.
- *Interoperability:* It send the speech features to cloud
- *Configurability:* It can be a configured remotely.

## BACKGROUND & RELATED WORKS

### A. Fog Computing & IoT - State of the Art

The Internet-of-Things (IoT) is defined as a framework that can interconnect sensors, actuators, and the cloud, communicating via the internet or other wireless networking capabilities such that end-users can benefit from connected intelligence [8]. Cisco coined the term "Fog Computing" [7] to describe a new category of devices and services that can provide computational intelligence at the edge of the networks. Fog computing is defined as a distributed computing paradigm that fundamentally functions as a middleman between the sensors and/or actuators and the cloud [8]. In other words, Fog computing generates three-tier architecture (clients ⟷ fog ⟷ cloud). The key benefits of adopting fog computing over traditional cloud computing include; reduction in network traffic [9] and data storage on the cloud; low response latency in IoT applications [5, 6, 10].

### B. Fog Computing in Healthcare

Fog computing is increasingly penetrating the area of healthcare, especially to improve telehealth and telemedicine infrastructure that promise to cope up with the rising healthcare needs in elderly population and individuals with chronic conditions around the world. For example, fog computing upgrades the standards of body sensor networks for medical signal processing [10], energy-efficient computing, and privacy and se-curity [9]. Recently, we developed EchoWear [3-6] that leverages an Android smartwatch to manage and tele-monitor in-home speech exercises of patients with PD. For example, each exercise session lasts for 15-20 minutes, generating 40-50 MB of speech data. In this paper, we developed FIT, a fog-computing interface to reduce the data complexity and introduce computational intelligence at the edge, rather than in the cloud. In the subsequent sections, we will unfold the architecture and functions of FIT.

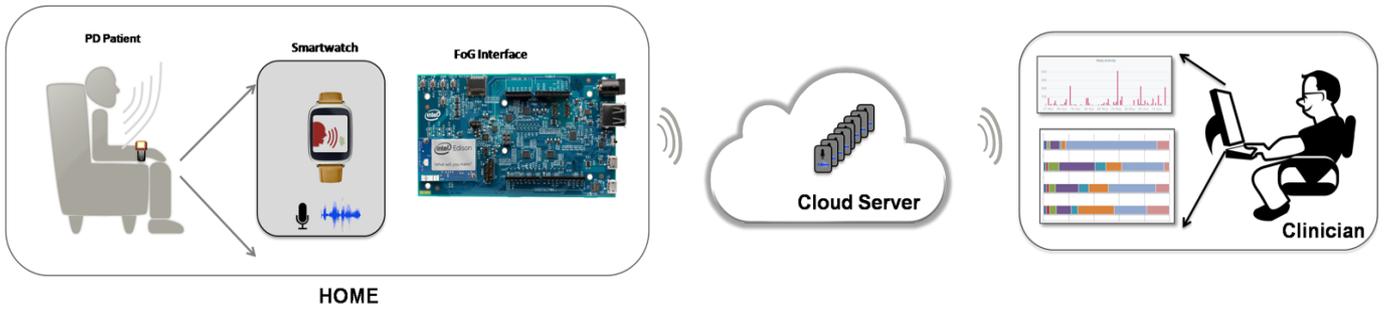

Fig. 2. The workflow of proposed Fog-driven IoT interface (FIT). The smartwatch acquired the audio data while patients with Parkinson's disease (PD) performed speech exercises. FIT acts as a smart gateway for processing clinical speech. The extracted speech features were synced to the cloud. FIT can be configured remotely via internet.

## FIT - FOG-DRIVEN IOT INTERFACE

The proposed fog-driven IoT interface, named as FIT is shown in Fig 2. It translates cloud-based speech analysis to fog computer based speech analysis. It consists of following components:

- <u>Smartwatch:</u> EchoWear app installed on android smartwatch facilitates the acquisition of clinical speech data.
- <u>Fog Computing Platform (Intel Edison):</u> The tablet receives the data from smartwatch and forward it to the fog computer that is a local node building computational and communication intelligence. It extracts clinical features from speech data.
- <u>Cloud:</u> Finally, the extracted features are sent to the secured cloud storage for long-term analysis. These databases can be queried by clinicians to tele-monitor the progress of their patients.

### A. Intel Edison as a Fog Interface

The Intel Edison was chosen because of its size and configurability capabilities. The essential components of the Intel Edison Board are that it supports a System on a Chip 22-nm Intel SoC that includes a dual-core, dual-threaded Intel Atom CPU at 500-MHz and a 32-bit Intel Quark micro-controller at 100 MHz. It has a random access memory of 1 GB LPDDR3 POP memory (2 channel 32bits @ 800MT/sec), a flash storage of 4 GB eMMC (v4.51 spec), wireless Broadcom 43340 802.11 a/b/g/n; Dual- band (2.4 and 5 GHz) with the option for on board antenna or external antenna [16]. Its compactness brings the wireless and Bluetooth 4.0 hardware. Another reason for using Edison is that we were able to increase its storage. Its expansion slot can be attached to an Arduino breakout kit board. The dimensions of the Edison is 60mm x 29mm x 8mm (the mini breakout board is attached with it). Intel Edison is a low power embedded board with power consumption of 3.3V - 4.5V @ 1W [15].

## EXPERIMENTS

We used python on the Intel Edison to process the audio files. The Edison is a low-power embedded computer with a Linux OS relying on Yocto so that we can compile C/C++ files, be able to run Python scripts, code in Node.js, and many other capabilities[14, 15]. The Secure Copy Protocol (SCP) was used for transfer of files and directories between two hosts (local or remote). Once the files were transferred over to Edison by android tablet, our python code was able to search for all of the files in a specific folder. Once the files were detected, it processed each file for getting the corresponding clinical features, file duration, and size. The speech features were extract-ed from audio files using programs written in Python.

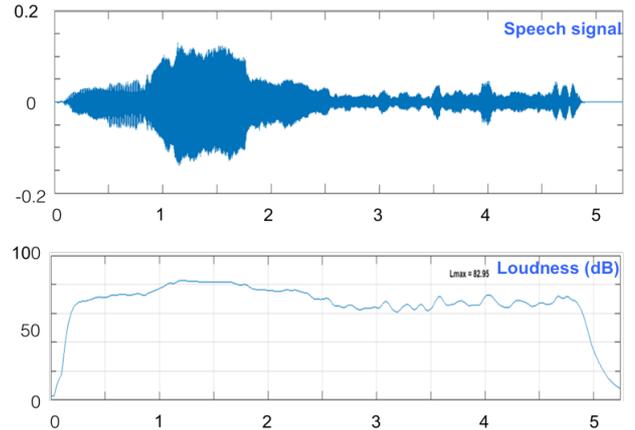

Fig. 3. Top sub-figure shows a speech signal and bottom sub-figure shows corresponding loudness in dB.

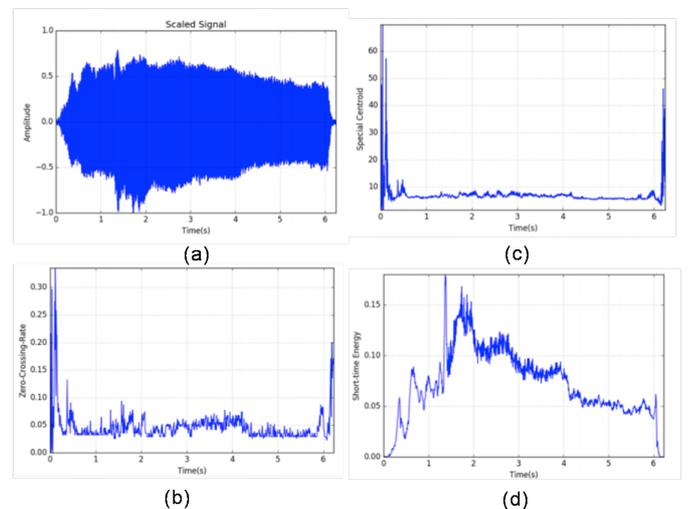

Fig. 4. (a) Speech signal; (b) Spectral centroid; (c) Zero-crossing rate; (d) Short-time energy

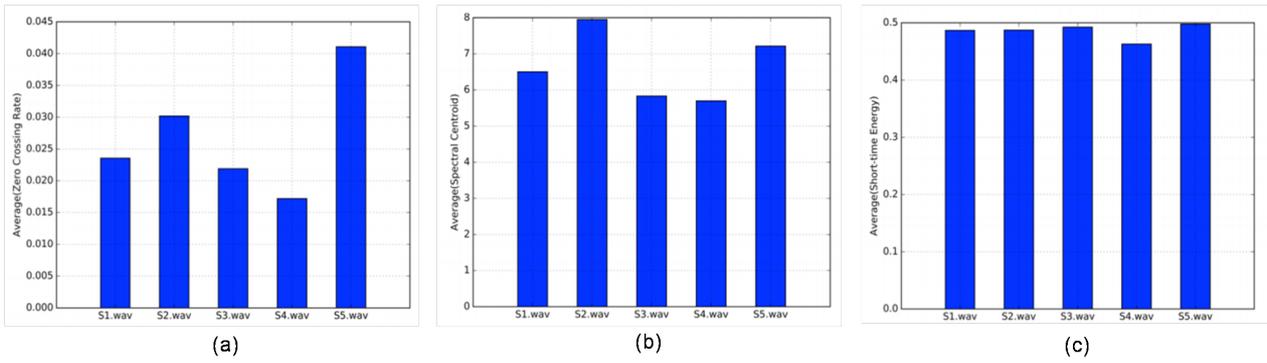

Fig. 5. (a) Average ZCR; (b) Average SC; (c) Average short-time energy for 5 speech files ($S_1$, $S_2$, $S_3$, $S_4$, and $S_5$).

## V. RESULTS & DISCUSSIONS

We conducted a month-long in-home trial involving six patients with PD [3]. The study involved commonly used speech tasks; sustained vowel phonation, low and high pitch range [4]. Following acoustic features were extracted from speech signal:

- *Perceptual Loudness:* It is the perceived intensity of the clinical speech signal. It is computed using various auditory models. We used the Zwicker's model [13] for loudness computation.
- *Zero-Crossing Rate:* It is used to analyze and recognize voiced-unvoiced parts of the speech. It is the number of times the signal magnitude crosses the value of zero [11].
- *Spectral Centroid:* It is center of gravity, the frequency bin where most of the speech energy is concentrated. It is a measure of 'brightness' of a sound [12].
- *Short-time Energy:* It reflects amplitude variations in the speech signal.

Fig. 3 shows dependence of loudness on the amplitude of speech signal. The instantaneous loudness is computed by taking overlapping time-windows of the speech signal. Fig. 4 shows the contour of speech features derived from a speech segment of length 6 second. It is evident that variations in speech signal is captured by these features. Five speech files ($S_1$, $S_2$, $S_3$, $S_4$, and $S_5$) were processed by Edison to compute the zero crossing rate (ZCR), spectral centroid (SC), and short-time energy. Table 1 shows time taken and size of each file. Each file took about a little more than 2 seconds to complete all computations (See Table 1). This shows quick processing of a clinical audio file and storage of features on the Fog computer. Fig. 5 shows the averages of these features that were computed by the Intel Edison for each of the audio file.

Table 1: Benchmarking the fog computer for computing the clinical speech features: zero crossing rate, special centroid, and short-time energy.

| Speech Tasks | Processing Time(s) | File Duration(s) | Size (kB) |
|---|---|---|---|
| S1 | 2.34 | 6.24 | 551 |
| S2 | 2.33 | 6.18 | 545 |
| S3 | 2.12 | 5.62 | 496 |
| S4 | 2.28 | 6.08 | 537 |
| S5 | 1.86 | 4.96 | 438 |

## VI. CONCLUSIONS

In this paper, we introduce a Fog computing architecture, FIT, for the analyses of speech data in remote settings such as homes. Intel Edison was used as a Fog computer to extract clinically-relevant features that are finally sent to the cloud. We demonstrated several functionalities of the FIT framework including configurability, computational intelligence, and interoperability. The results showed that FIT is a useful interface for remote monitoring of speech treatments.

## REFERENCES

[1] Dorsey ER et al. Projected number of people with Parkinson disease in the most populous nations, 2005 through 2030. Neurology 2007;68:384-6.

[2] Ho A, Iansek R, Marigliani C, Bradshaw JL, Gates S. Speech impairment in a large sample of people with Parkinson's disease. Behav Neurol 1998; 11:131-137.

[3] Mahler, L; Dubey, H.; Goldberg,C.; and Mankodiya, K., Use of Smartwatch Technology for People with Dysarthria, Motor Speech Conference, Madonna Rehabilitation Hospital, March 2016.

[4] Dubey, H., et al. "EchoWear: smartwatch technology for voice and speech treatments of patients with Parkinson's disease." *Proceedings of the conference on Wireless Health*. ACM, 2015.

[5] Dubey, H., et al. "A multi-smartwatch system for assessing speech characteristics of people with dysarthria in group settings."*Proceedings e-Health Networking, Applications and Services (Healthcom), 2015 IEEE 17th International Conference on, Boston, USA*. 2015.

[6] Dubey, H., et al. "Fog Data: Enhancing Telehealth Big Data Through Fog Computing." *Proceedings of the ASE Big Data & Social Informatics 2015*. ACM, 2015.

[7] F. Bonomi, R. Milito, J. Zhu, and S. Addepalli, "Fog computing and its role in the internet of things," in Proceedings of the first edition of the MCC workshop on Mobile cloud computing. ACM, 2012, pp. 13–16.

[8] Dastjerdi, Amir Vahid, et al. "Fog Computing: Principals, Architectures, and Applications." *arXiv preprint arXiv:1601.02752* (2016).

[9] Luan, T.H., Gao, L., Li, Z., Xiang, Y., We, G. and Sun, L., 2016. A View of Fog Computing from Networking Perspective. *arXiv preprint arXiv:1602.01509*.

[10] Nguyen Gia, T et. al (2015). Fog Computing in Body Sensor Networks: An Energy Efficient Approach. In *IEEE International Body Sensor Networks Conference (BSN'15)*.

[11] Bachu, R. G., et al. "Separation of voiced and unvoiced using zero crossing rate and energy of the speech signal." A*merican Society for Engineering Education (ASEE) Zone Conference Proceedings*. 2008.

[12] Paliwal, Kuldip K. "Spectral subband centroid features for speech recognition." *Acoustics, Speech and Signal Processing, 1998. Proceedings of the 1998 IEEE International Conference on*. Vol. 2. IEEE, 1998.

[13] Zwicker, Eberhard, and Hugo Fastl. *Psychoacoustics: Facts and models*. Vol. 22. Springer Science & Business Media, 2013.

[14] Yocto Project, A Linux Foundation Collaborative Project, https://www.yoctoproject.org/question/what-yocto-project, online; accessed 2 January 2016.

[15] JIMBO, and HELLOTECHIE. "Edison Getting Started." *Sparkfun*. Sparkfun, 10 Mar. 2016. Web. https://learn.sparkfun.com/tutorials/edison-getting-started-guide.